\documentstyle[12pt]{article}
\begin{document}

\begin{center}
\bf
\LARGE
Analysis of Data on Low-Energy $\gamma p$ Scattering \\and 
Determination of Proton Polarizabilities\\
\end{center}
\begin{center}
\large
P.S.~Baranov, A.I.~L'vov, \underline{V.A.~Petrun'kin}, L.N.~Shtarkov\\
P.N.~Lebedev Physical Institute, Moscow, Russia\\
\end{center}
\begin{center}
\large
Abstract\\
\end{center}

For the first time an analysis of {\em all} experimental data on the
differential cross section of elastic $\gamma p$ scattering at photon
energies $\omega<150$ MeV is performed in order to determine the
electric $(\alpha_p)$ and magnetic $(\beta_p)$ polarizabilities of the
proton. A fit of these data with the two free parameters, $\alpha_p$ and
$\beta_p$, embedded into a theoretical cross section obtained on the
basis of finite-energy $s$-channel dispersion relations gives the
following world-average values of the proton polarizabilities (in units
of $10^{-4}$ fm$^3$):\\
$\alpha_p^{exp} = 11.7\pm0.8~(\rm stat.+sys.)\pm0.7~(model)$,
$\beta_p^{exp}=2.3\pm0.9~(\rm stat.+sys.)\pm0.7~(model)$,
where the first error takes into account statistical and systematic
errors of the experimental cross sections and the second error does
model uncertainties in the theoretical cross section.

\begin{center}
I. Introduction\\
\end{center}

Experimental studies of elastic $\gamma p$ scattering began in the
middle of 50's years [1-6], but for the first time the proton
polarizabilities $\alpha_p$ and $\beta_p$ were extracted from
experimental cross sections at photon energies $\omega \le110$ MeV in
the works [3,6]. Results of new experiments [7-10] on $\gamma p$
scattering and new determinations of the values of $\alpha_p$ and
$\beta_p$ were published only in 90's. However, statistical and
systematic errors in the experimental values of the proton
polarizabilities (especially in $\beta_p$ and $\alpha_p+\beta_p$) from
the works [7-10] are yet not small.

This talk presents some results of our analysis concerning a
compatibility of all data on elastic $\gamma p$ scattering at photon
energies $\omega<150$ MeV and the values of $\alpha_p$ and $\beta_p$
determined from different sets of the data.

\begin{center}
II. Analysis of data and results\\
\end{center}

All experimental data on the differential cross section of elastic
$\gamma p$ scattering at photon energies $\omega<150$ MeV from FIAN,
MAMI, SAL and other centres are split into two sets: 46 early data
points of 1955-1974 [1-6] and 48 recent points of 90's [7-10]. In our
analysis we take into account that authors of three works [5-7] made
corrections to their data afterwards. A compilation of these data, with
corrections, and a description of the calculation of the theoretical
cross section which we use for fitting the experimental data points are
given in works [11,12].

An important point of our analysis is a combined use of the statistical
and systematic experimental errors in a special way that was proposed
in our earlier work [13]. The function $\chi^2$ used for fitting
experimental cross sections is written as
\begin{equation}
\chi^2=\sum_{j=1}^{N^{exp}}\left\{\sum_{i=1}^{n_j}\left(\frac{k_j
\sigma_{ij}^{exp}-\sigma_{ij}^{th}(\alpha,\beta)}{k_j
\Delta_{ij}^{exp}}\right)^2 +\left(\frac{k_j-1}{k_j
\delta_j}\right)^2\right\}
\end{equation}
where $j$ is the experiment number, $i$ is the experimental point
number, $\sigma_{ij}^{exp}$ and $\sigma_{ij}^{th}(\alpha,\beta)$ are
the experimental and theoretical cross section at the photon energy
$\omega_i$ and the scattering angle $\theta_i$,
$\Delta\sigma_{ij}^{exp}$ is the statistical error of an individual
point, $\delta_j$ is the systematic error for the $j$-th experiment,
$k_j$ is a normalization factor to be found for the $j$-th experiment,
$\alpha$ and $\beta$ are the proton polarizabilities to be found.

The first step of our analysis was in separate fits of experimental
data from each of the quoted works. Results are shown in Table 1.
Since experimental data in the works [4,8] are taken at fixed angle,
two polarizabilities cannot be found and we fit those data using fixed
$\alpha_p + \beta_p = 14.0 \pm 0.5$ [12] calculated from a dispersion sum rule.

The confidence levels, shown in the Table 1, are sufficiently high 
($P\ge 12\%$). For this reason no experiment can be excluded from our 
following analysis. Below we assume the possibility to combine the data 
of different experiments [1-10] taking into account the errors and 
a spread in the found values of $\alpha_p$ and $\beta_p$.

Note that fitting of a subset of data from the work [10] obtained only with
the tagged photons leads to too large values of $\alpha_p$ and
$\beta_p$ and their errors (the last line in Table~1). Consequently,
experiments with the bremsstrahlung photon beam have a certain
advantage for a determination of $\alpha_p$ and $\beta_p$ as compared
with tagged photon experiments, at least at the present level of
statistical errors in the differential cross section of elastic
$\gamma p$ scattering.

The second step of the analysis was fitting two sets, all the early 46
points and all the recent 48 points. After that, a fit was done of all
94 points. The corresponding global-average polarizabilities are shown
in Table 2.  One can see that there is a satisfactory agreement between
the early and recent global averages at the existing level of errors.
Consequently, there are no reasons to ignore the early experiments, and
it is better to combine all the experimental points.  The averaged
values of the proton polarizabilities show an appreciable decrease of
their errors in comparison with individual errors given in the Table 1.

The third step of the analysis was in a variation of model parameters
in the theoretical cross section obtained on the basis of finite-energy
$s$-channel dispersion relations and in estimation of model
errors in the extracted values of $\alpha_p$ and $\beta_p$.  The most
important contributions to the model errors at $\omega<150$ MeV (and at
$\omega<100$ Mev) are shown in Table 3.  Summing these contributions in
quadrature we get approximate total model errors
$$
\Delta\alpha_p^{mod}\approx\Delta\beta_p^{mod}\approx0.7 ~~~\rm at~
\omega<150~MeV
$$
and
$$
\Delta\alpha_p^{mod}\approx\Delta\beta_p^{mod}\approx0.3~~~ \rm at~
\omega<100~MeV.
$$
Finally, taking results from Table 2 (for all data) and
adding the model errors we get the world-average values of the proton
polarizabilities:
\begin{equation}
\alpha_p^{exp}=11.7\pm0.8~(\rm stat.+syst.)\pm0.7~(model),
\end{equation}
\begin{equation}
\beta_p^{exp}=2.3\pm0.9~(\rm stat.+syst.)\pm0.7~(model).
\end{equation}
It is seen that the experimental and model errors are comparable.  The
values (2) and (3) may be recommended for a general use
and for PDG publications.
Note that the current PDG values of the proton polarizabilities are
based on results of only 4 recent experiments [7-10] and, moreover, they
are obtained at the fixed sum $(\alpha_p+\beta_p)^{th}=14.2\pm0.5$ in
order to lower real experimental errors.

In conclusion we would like to emphasize that in order to reduce the
experimental and model errors in the extracted values of $\alpha_p$ and
$\beta_p$ it is necessary to make new and more complete measurements of
the differential cross section of $\gamma p$ scattering with the
statistical and systematic errors $\sim1{-}2\%$ [14] in the energy region
$\omega<100$ MeV where the model errors are essentially lower.

This work was supported by
RFFI grants 96-02-17590 and 97-02-71019.
%%\vspace{2cm}
\begin{center}
\large
References
\end{center}
1. C.~Oxley, Phys.~Rev.~106 (1958) 733; C.~Oxley, V.~Telegdi,
Phys.~Rev.~100 (1955) 435.\\
%%\ \\
2. L.~Hyman et al., Phys.~Rev.~Lett.~3 (1959) 93.\\
%%\ \\
3. V.I.~Goldansky et al., Nucl.~Phys.~18 (1960) 473.\\
%%\ \\
4. G.~Bernardini et al., Nuovo Cim.~18 (1961) 1203.\\
%%\ \\
5. G.~Pugh et al., Phys.~Rev.~105 (1957) 982; MIT Summer Study (1967) 555.
   \par
D.~Frish, private communication (1967).\\
%%\ \\
6. P.S.~Baranov et al., Phys.~Lett.~B52 (1974) 122.\\
%%\ \\
7. F.J.~Federspiel et al., Phys.~Rev.~Lett.~67 (1991) 1511.
    \par
   A.M.~Nathan, private communication (1993).\\
%%\ \\
8. A.~Zieger et al., Phys.~Lett.~B278 (1992) 34.\\
%%\ \\
9. E.L.~Hallin et al., Phys.~Rev.~C48 (1993) 1497.\\
%%\ \\
10. B.E.~MacGibbon et al., Phys.~Rev.~C52 (1995) 2097.\\
%%\ \\
11. P.S.~Baranov et al., preprint FIAN 50, Moscow (1999) (in Russian);
\par
Fiz.~Elem.~Chast.~Atom.~Yad., vol.32, issue 3 (2001) (in press).\\
%%\ \\
12. A.I.~L'vov et al., Phys.~Rev.~C55 (1997) 359.\\
%%\ \\
13. P.S.~Baranov et al., Yad.~Fiz.~5 (1967) 1221 (in Russian).\\
%%\ \\
14. D.V.~Balin et al., preprint 2104, Petersburg Nuclear Physics
Institute (1996).

\vspace{1cm}
Table 1. Polarizabilities of the proton in units of $10^{-4}$ fm$^{3}$
extracted from data of different experiments [1-10]. $n$ is the number
of experimental points. Uncertainties take into account both
statistical and systematic errors of the experimental cross sections of
$\gamma p$ scattering. $P$ is the confidence level for the shown
$\chi^2$.  The last line shows values extracted from a subset of
data of [10] obtained only with the tagged photons.

\vspace{0.5cm}
\begin{tabular}{|l|r|r|r|r|r|r|}
\hline
Experiment &  $n$ &  $\alpha_p$~~~~ &  $\beta_p$~~~~ &  $\alpha_p+\beta_p$~
&  $\chi^2/N_f$ &  $P(\%)$\\ \hline
\hline
%\hline
Oxl58~[1]&  4 &17.0$\pm$8.1 &  $-$6.7$\pm$3.7 &  10.2$\pm$9.2 &  4.2/2~
&  12\\
Hym59~[2]&  12 &13.9$\pm$5.6 & $-$4.7$\pm$7.2 &  9.2$\pm$6.1 &  0.6/10
&  100\\
Gol60~[3]&  5 &10.1$\pm$7.8 &  9.0$\pm$5.0 &  19.1$\pm$10.2 &  2.3/3~
&  52\\
Ber61~[4]&  2 &11.4$\pm$2.9 &  2.6$\pm$2.9 & --~~~~  &  0.7/1~ &  41\\
Fri67~[5]&  16 &14.2$\pm$4.0 &  5.6$\pm$4.2 &  19.8$\pm$4.3 &  2.4/14
&  100\\
Bar74~[6]&  7 &11.4$\pm$1.4 &  $-$4.7$\pm$2.5 &  6.7$\pm$3.3 &  8.0/5~
&  15\\
Fed91~[7]&  16 &13.7$\pm$3.7 &  2.1$\pm$3.1 &  15.9$\pm$4.4 &  17.3/14
&  24\\
Zie92~[8]&  2 &10.0$\pm$1.4 &  4.0$\pm$1.4 & --~~~~  &  0.1/1~ &  73\\
Hal93~[9]&  12 &9.1$\pm$1.7 &  3.7$\pm$1.5 &  12.7$\pm$2.0 &  5.9/10
&  82\\
Mac95~[10]&  18 &12.2$\pm$1.7 &  3.3$\pm$1.8 &  15.5$\pm$3.1 &  7.4/16
&  97\\
\hline
Mac95(tagged)~[10]&  8 &18.3$\pm$5.7 &  13.2$\pm$7.2 &  31.5$\pm$12.3
&  2.2/6~ &  90\\
\hline
\end{tabular}

\newpage
Table 2. The global-average proton polarizabilities in units of
$10^{-4}$ fm$^{3}$ extracted from early data [1-6], from recent data
[7-10], and from all the data. Errors take into account statistical and
systematic errors of the experimental cross sections.

\vspace{0.5cm}
\begin{tabular}{|l|c|c|r|c|r|c|}
\hline
Exp. & $n$ & $\alpha_p$ & $\beta_p$~~~~ & $ \alpha_p$+$\beta_p$
&  $\alpha_p-\beta_p$~ &  $\chi^2/N_f$ \\ \hline
\hline
50-70's~[1-6]&  46 &12.8$\pm$1.1 &  $-$0.3$\pm$1.6 &  12.5$\pm$2.2
&  13.0$\pm$1.7 &  33.1/44\\
90's~[7-10]&  48 &10.8$\pm1.0$ &  3.2$\pm1.0$ &  14.0$\pm$1.6 &  7.7$\pm$1.2
&  33.7/46\\
all~[1-10]&  94 &11.7$\pm$0.8 &  2.3$\pm$0.9 &  14.0$\pm$1.3 &  9.5$\pm$1.0
&  73.1/92\\
\hline
\end{tabular}

%%\newpage
\vspace{1cm}
Table 3. Uncertainties in the values of extracted polarizabilities (in
units of $10^{-4}$ fm$^3$) related with a model dependence of the
theoretical cross section.  Changes shown in the extracted $\alpha_p$
and $\beta_p$ emerge from using different values of pion
photoproduction amplitudes near threshold (SAID and HDT), from
increasing the value of the resonance pion-photoproduction amplitude
$M_{1+}$, from ignoring contributions of pion-pair photoproduction in
dispersion integrals, from changing other parameters of the dispersion
theory:  $M_1$ (``$\sigma$-meson mass"), $g_{\pi NN}$,
$\gamma_{\pi}^{(non-\pi^{\circ})}$. Total model uncertainties are
shown in the last line. The model uncertainties shown in parentheses
are related to the case when only experimental data at $\omega<100$ MeV
are used.

\vspace{0.5cm}
\begin{tabular}{|l|c|c|c|c|}
\hline
Changes & $\delta\alpha_p$ & $\delta\beta_p$
& $\delta(\alpha_p+\beta_p)$ & $\delta(\alpha_p-\beta_p)$ \\ \hline
\hline
SAID $\rightarrow$ HDT & $-$0.44 & $-$0.08 & $-$0.51 & $-$0.36 \\
 & ($-$0.19) & (+0.03)& ($-$0.16) & ($-$0.22) \\
\hline
$M_{1+}\rightarrow +2\%$ & +0.18 & $-$0.11 & +0.06 & +0.29\\
 & (+0.07)& ($-$0.04) & (+0.03)& (+0.11)\\
\hline
without $\pi\pi$ photoproduction& +0.00 & $-$0.09 & $-$0.09 & +0.09\\
 & (+0.00)& ($-$0.02) & ($-$0.02) & (+0.02)\\
\hline
$M_1=500$ MeV $\rightarrow$ 700 MeV & $-$0.35 & +0.53 & +0.19 & $-$0.88 \\
 & ($-$0.20) & (+0.24)& (+0.04)& ($-$0.44) \\
\hline
$\mid g_{\pi NN} F_{\pi\gamma\gamma}\mid$ $\rightarrow +4\%$ & $-$0.14
& +0.10 & $-$0.04 & $-$0.24\\
& ($-$0.09)& (+0.07)& ($-$0.02)& ($-$0.16)\\
\hline
$\gamma_{\pi}^{(non-\pi^{\circ})}=5.5\rightarrow$ 7.3 & +0.40 & $-$0.45
& $-$0.05 & +0.84\\
& (+0.18)& ($-$0.15)& (+0.03)& (+0.32)\\
\hline
\hline
Total model uncertainty & ~~0.72 & ~~0.72 & ~~0.56 & ~~1.33\\
& ~~(0.34)& ~~(0.29)& ~~(0.17)& ~~(0.62)\\
\hline
\end{tabular}
\end{document}